\documentclass{PoS}
\usepackage{graphicx}
\usepackage{epsfig}
\usepackage{amssymb}

\title{Fluctuations, correlations and the sign problem in QCD}

\ShortTitle{Fluctuations, correlations and the sign problem in QCD}

\author{\speaker{M. P. Lombardo} \\
        INFN Laboratori Nazionali di Frascati, I-00044, Frascati (RM), Italy\\
        E-mail: \email{lombardo@lnf.infn.it}}

\author{K. Splittorff 
\\The Niels Bohr Institute, Blegdamsvej 17, DK-2100, Copenhagen
  {\O}, Denmark
\\ E-mail: \email{split@nbi.dk}}

\author{J.J.M. Verbaarschot
\\Department of Physics and Astronomy, SUNY, Stony Brook,
 New York 11794, USA
\\E-mail: \email{jv@chi.physics.sunysb.edu}}

\abstract{We study the distribution of the phase angle and the magnitude of the
fermion determinant as well as its correlation with the chiral condensate and the
baryon number for QCD at non-zero quark chemical potential. Results are derived to one-loop order
in Chiral Perturbation Theory (ChPT), as well as by analytical and numerical calculations in QCD
in one Euclidean dimension. We find a qualitative change of the
distribution of the phase of the fermion determinant when the quark
mass enters 
the spectrum of the Dirac operator: it changes from a periodicized
Gaussian distribution to a periodicized Lorentzian distribution. We
also explore the possibility that some observables 
remain weakly correlated with the phase of the fermion determinant even though
the sign problem is severe. We discuss the practical implications of
our findings on lattice simulations of QCD at non-zero baryon chemical 
potential.}

\FullConference{The XXVII International Symposium on Lattice Field Theory\\
		 July 26-31, 2009\\
		 Peking University, Beijing, China}

\begin{document}

\section{Introduction}

It is well known that standard lattice simulations based on importance sampling
are not possible for QCD at non-zero chemical potential, since the measure in
the functional integral is no longer positive definite. In the past years
a few methods have been proposed to circumvent this problem, at least for
small values of $\mu/T$ \cite{rev09}. They
all share a common philosophy: choose wisely a simulation ensemble -
either at zero or non-zero imaginary chemical potential  
and fluctuations will allow the exploration of the target ensemble at
non-zero baryon density. In a natural way, we are thus
led to consider the overlap between the simulation ensemble and
the target ensemble, and to devise strategies to maximize it.
 This note, which is
based on a more extended publication \cite{lsv1}, reports our first results
towards this goal. 

Fluctuations increase with temperature    
so in general ensembles generated a higher temperature
allow a better re-weighting \cite{Fodor:2001pe}, 
an easier calculation of
the Taylor coefficients \cite{bf},  or a safer extrapolation from
imaginary chemical potential \cite{im1,im2,im3,im4}. Further,  the shape of the distribution
function plays a major role in the density of states method \cite{dos1,dos2}
and some assumptions on its shape are needed. 

The main goal of our investigation is to put these heuristic considerations
on quantitative grounds by a combination of analytic studies
and numerical simulations  of chiral perturbation theory and QCD
  in one dimension.

We will base our discussion mostly on 
the $\theta$-distribution 
$\langle\delta(\theta-\theta')\rangle_{N_f}$ and the 
constrained distribution 
$
 \langle \cal{O} \ \delta(\theta-\theta')\rangle$ (the phase of the fermion determinant is denoted by $\theta'$).
The $\theta$-distribution will help
 assessing the  overlap between the simulation and target ensembles, and
the  constrained distribution
shows how averages are built up in the spirit of the density of states method
(the integral over $\theta$ obviously gives the full expectation value
 $\langle {\cal O}\rangle$).  The distribution of an
observable over the phase allows us to address which range of the phase is
essential for the full expectation value of $\cal{O}$.

\section {Results from Chiral Perturbation Theory}
Let us first recall that the $\theta$ distribution in the full theory
\begin{equation}
\langle\delta(\theta-\theta')\rangle_{N_f}d\theta
=\frac{\int dA |\det(D+\mu\gamma_0+m)|^{N_f} e^{iN_f\theta'}
  \delta(\theta-\theta') e^{-S_{\rm YM}}}
{\int dA |\det(D+\mu\gamma_0+m)|^{N_f} e^{iN_f\theta'} e^{-S_{\rm YM}}} d\theta \end{equation}
has a simple expression in terms of the phase quenched distribution
\begin{equation}
 \langle \delta(\theta-\theta')\rangle_{N_f} = e^{i\theta N_f}\frac{Z_{|N_f|}}{Z_{N_f}}\langle \delta(\theta-\theta')\rangle_{|N_f|}.  
\end{equation}
In the following, we will mostly consider the two flavor case
($Z_{1+1^*} \equiv Z_{|2|}$ and $Z_{1+1} \equiv Z_2$):
\begin{equation}
 \langle \delta(\theta-\theta')\rangle_{1+1} = e^{2i\theta}\frac{Z_{1+1^*}}
{Z_{1+1}}\langle \delta(\theta-\theta')\rangle_{1+1^*}.  
\end{equation}

The calculations follow
the general method of \cite{method}: the $\theta$ distribution is 
obtained from the moments of the phase factor 
\begin{equation}
\langle\delta(\theta-\theta')\rangle_{N_f}
=\frac{1}{2\pi} \sum_{p=-\infty}^\infty e^{-ip\theta}\langle
e^{ip\theta'}\rangle_{N_f}. 
\end{equation}

First we consider $\mu<m_\pi/2$ where the leading order
  difference between the phase quenched and full free energies is
  determined by the one-loop corrections.
 The $\theta$ distribution to this order is   
given by a periodicized Gaussian
\begin{equation}
\hspace{1cm}
\langle\delta(\theta-\theta')\rangle_{1+1} = \frac{e^{2i\theta}}{\sqrt{\pi
    \Delta G_0}}{\sum_{n=-\infty}^{\infty}e^{-(\theta+2\pi n)^2/\Delta G_0+\Delta G_0}.}
\label{th-dist}
\end{equation}
where 
$G_0 $ is the free energy difference between neutral and charged pions.
The distribution of the baryon number $n_B$ over $\theta$ is given by
\begin{equation}
\langle n_B \ \delta(\theta-\theta')\rangle_{1+1} 
 =  \left(\lim_{\tilde{\mu}\to\mu} \frac{d}{d\tilde{\mu}} \Delta
   G_0(-\mu,\tilde{\mu})\right)\sum_{n=-\infty}^{\infty}(1+i\frac{\theta+2\pi
   n}{\Delta G_0})\frac{e^{2i\theta}}{\sqrt{\pi \Delta
    G_0}}e^{-(\theta+2\pi n)^2/\Delta G_0+\Delta G_0}.
\end{equation}
Note that, since 
$\langle n_B \rangle = 0 $  in ChPT,  the integral of the above distribution is zero as it should  In contrast, the integral of
  $\theta$-distribution is unity. 
The similarity of the two distributions displayed in  Fig. 1
clearly shows that the details of the distribution are important 
to get the correct results for $n_B$, especially when $G_0$ is large,
i.e. when the sign problem is severe. 
\begin{figure}[t]
\vspace*{-5truecm}
  \unitlength1.0cm
\hspace*{-1.5 truecm} \epsfig{file=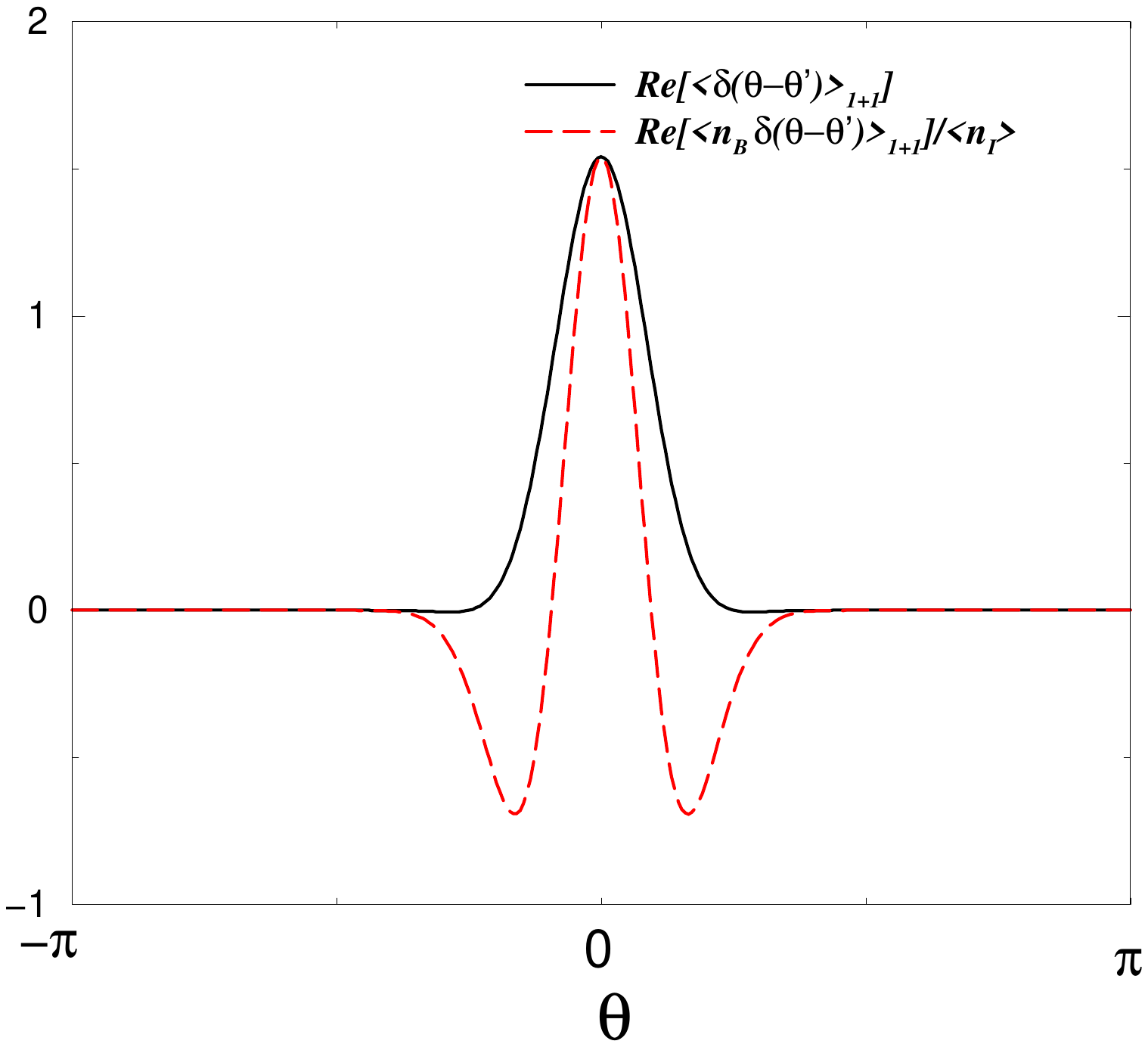,width=9cm} \hspace{-2mm}
 \epsfig{file=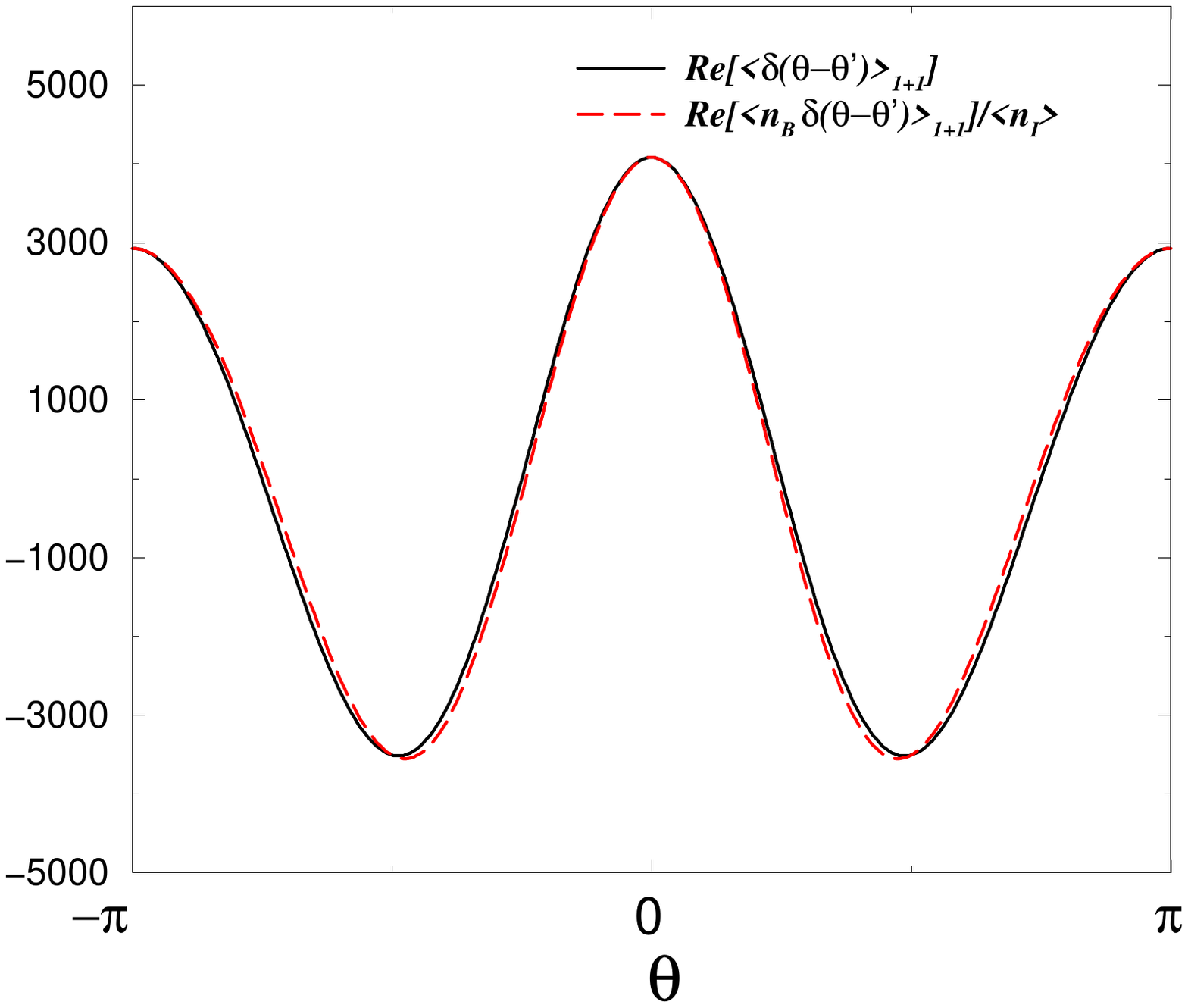,width=9.0cm}
  \caption{\label{fig:nBth-dist} The real part of the distribution of the
    phase $\langle\delta(\theta-\theta')\rangle_{1+1}$ (solid curve) 
    for a small $\Delta G_0=0.2$ (left) and a large $\Delta G_0=10$ value 
(right).    Also shown is the real
    part of the distribution of the 
    baryon number over $\theta$ (dashed curve). 
In either cases the $\theta$ distribution is normalized to 1, while
the integral of distribution of the baryon over $\theta$  is zero.
    This directly illustrates the severity of the sign problem at large $G_0$} 
\end{figure}

In order  to make contact between these calculations and
lattice studies, we have also have computed the phase quenched as
  well as the partially quenched $\theta$ distribution where the
  ensemble is generated at zero chemical potential \cite{Ejiri}.
The relationship between the $\theta$ distributions derived in one-loop
ChPT is
surprisingly simple:  Whether we compute the width of the Gaussian
for the $\theta$-distribution in the full ensemble generated at $\mu$, or the
partially quenched ensemble generated at $\mu=0$, or in the quenched ensemble, or the 
phase quenched ensemble, we find exactly the same result.
These results directly apply to the interpretation of lattice studies
at small chemical potential \cite{Ejiri,Ejiri1}.

\newpage

We now proceed to the computation of the $\theta$-distribution for $\mu>m_\pi/2$. Only the
even moments are accessible in ChPT and give the distribution   
\begin{equation}
\langle \delta(2\theta -2\theta')\rangle_{N_f} =
\frac 1{\pi} \sum_{p=-\infty}^\infty e^{-2ip\theta} 
\langle e^{2ip\theta'} \rangle_{N_f}.
\end{equation}
The leading order contribution now enters already at mean field level. 
This leads to the result 
\begin{equation}
\langle\delta(2\theta-2\theta')\rangle_{1+1}=
e^{2i\theta}
\frac{e^{V L_B}}{\pi} \frac{\sinh(V L_B)}{\cosh(V L_B)-\cos(2\theta)},
\end{equation}
where $L_B$ is the difference of the free energy densities in the full
and the phase quenched theory.
The quenched result is obtained simply removing the general factor
$e^{2i\theta}e^{V L_B}$. In both cases the distribution is a
periodicized Lorentzian. This is in sharp contrast 
with Gaussian obtained at low $\mu$, and implies that
the hypothesis leading to the central limit theorem are not realized
for $\mu>m_\pi/2$.

Further, we discuss the 
the distribution of
 \begin{equation}
F=|\det(D+\mu\gamma_0+m)|/\det(D+m) \equiv \exp(f). 
\end{equation}
This quantity was studied in lattice QCD \cite{Ejiri}
using the Taylor expansion method.
We have verified the assumption \cite{Ejiri}  that
the $\theta$ distribution remains Gaussian even for a fixed value of
$F$. Of course this applies only to $\mu < m_\pi/2$.

Since determinants fluctuate by many orders of
magnitude we feel that it is more appropriate to analyze the distribution
of $\log F \equiv f$ instead. The two distributions are related by a simple
transformation
\begin{equation}
\langle \delta(f-f')\rangle = F \langle \delta(F-F')\rangle.
\end{equation}
The  fluctuations of $f$ are induced by
the gauge field fluctuations.

We find that for
$\mu<m_\pi/2$ to one-loop order in ChPT 
the distribution of $f$ is Gaussian
\begin{equation}
\langle\delta(f-f')\rangle_{N_f} = \frac 1{\sigma_f \sqrt {2\pi}} 
e^{-\frac{(f-N_fE_f/2)^2}{2\sigma_f^2}}.
\end{equation}
where we have given the result for an arbitrary number of flavors $N_f$.
Note that $E_f$ and $\sigma_f$ do not depend on $N_f$, see \cite{lsv1}.

\section{QCD in one Euclidean dimension}
Consider now  one-dimensional QCD with a massive staggered fermion and gauge group $U(N_c)$  
 \begin{equation}
Z_{N_f}(\mu_c,\mu) = \int_{U(N_c)} dU \det M
\end{equation}
where 
\begin{equation}
\det M =  2^{-nN_c}\det[e^{n\mu_c} + e^{-n\mu_c}+ e^{n \mu} U +e^{-n \mu}
U^\dagger], \qquad U\in U(N_c)
\end{equation}
and $dU$ is the Haar measure. The group integral can be solved analytically
\cite{oned1,oned2}, which shows explicitly that the 
partition function does not depend on $\mu$, as expected since there
are no baryons in a $U(N_c)$ model. However, the quenched
model has a transition to a pion condensed phase at $\mu = \mu_c$.
So it resembles ordinary QCD with $\mu < \mu_B/3$, in contrast 
to $SU(N_c)$ in one dimension 
where the transition of the quenched and the full model
are very close to each other (which is of course the reason why we decided
to work in $U(3)$ rather than in $SU(3)$). 

Again the $\theta$ distribution follows form the moments
\begin{eqnarray}
\langle e^{2ip\theta'} \rangle
&=& \int_{U(N_c)} dU \frac { {\det}^p M}{ {\det}^p M^\dagger}.
\label{mom-1d}
\end{eqnarray}

For $\mu < m_\pi/2$ we obtain
\begin{equation} 
\langle e^{2ip\theta'} \rangle = \langle e^{2i\theta'} \rangle^{p^2} = \left(1
  -  \frac {\mu^2}{\mu_c^2}\right)^{p^2}. 
\end{equation}
With $\Omega\equiv -\log(1-\mu^2/\mu_c^2)$ this yields the
  quenched distribution
\begin{equation}
\langle\delta(\theta-\theta')\rangle = \frac{1}{\sqrt{\pi
    \Omega}}e^{-\theta^2/\Omega}
\qquad {\rm for} \qquad     \mu<\mu_c, \  N_c\to \infty,
\end{equation} 
valid for $2\theta \in[-\infty,\infty]$.

For $\mu > \mu_c$ the even moments  can be computed exactly for any 
  $N_c$. They are given by
\begin{equation}
\langle e^{2ip\theta'} \rangle  =e^{-2n|p|N_c \mu}, \qquad
\label{power}
\end{equation}
and the quenched $\theta$ distribution again becomes a periodicized Lorentzian
\begin{equation}
\langle\delta(2\theta-2\theta')\rangle 
= \frac{1}{\pi}\frac{\sinh(2nN_c \mu)}{\cosh(2nN_c \mu)-\cos(2\theta)} \qquad {\rm for} \qquad 
\mu>\mu_c \ , \ 2\theta \in[-\pi,\pi]. 
\end{equation}
The results have been verified numerically with a good accuracy, see
Figure 2.
\begin{figure}
\mbox{\epsfig{file=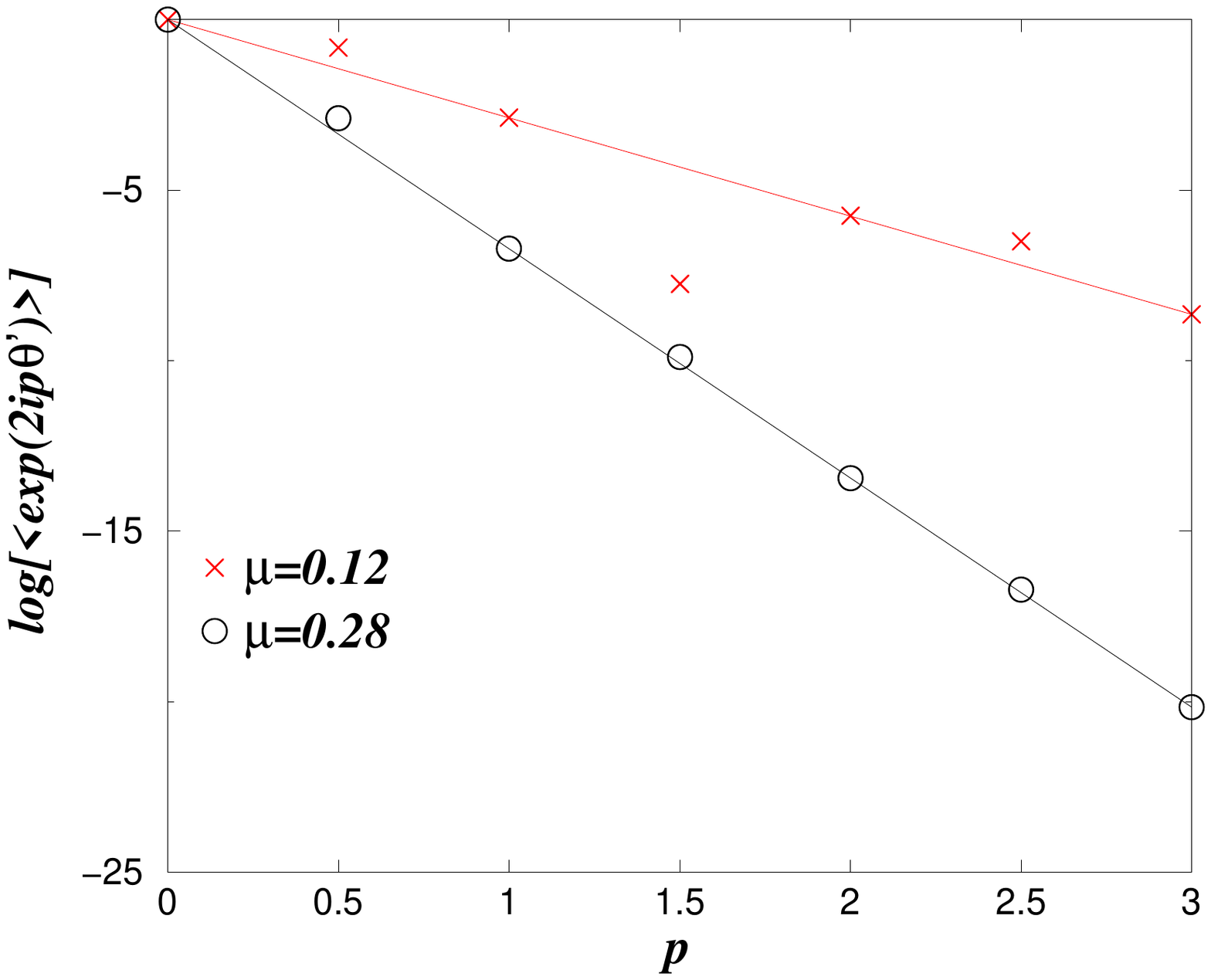, width=6.3cm,bb=0 40 600 430,clip=}}
\mbox{\epsfig{file=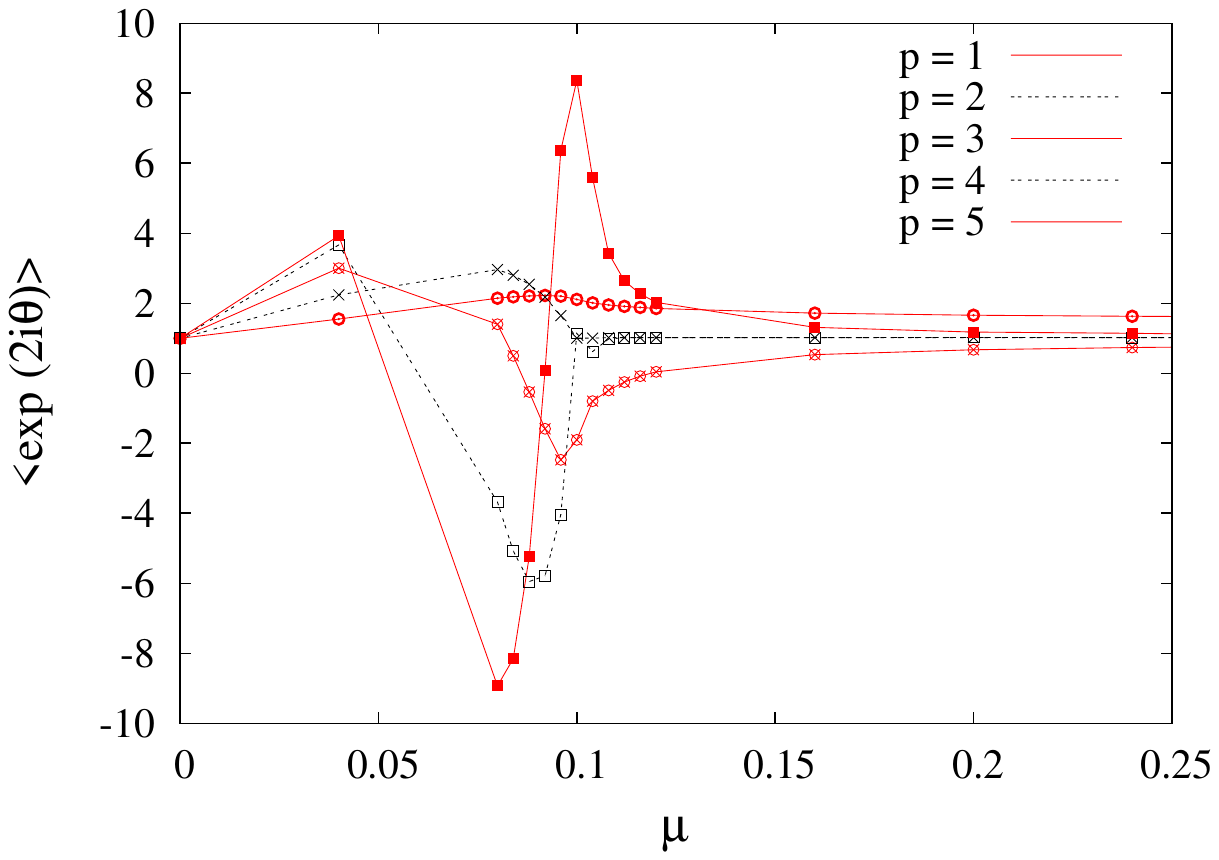,width=7cm,bb=0 50 400 300,clip=}}
\caption{$U(3)$ lattice QCD in one dimension for  $m_q =0.1$ i.e. $\mu_c = \sinh^{-1} m_q$.   
Left we show  the logarithm
of the $p-$th moment as a function of $p$, for $\mu = (0.12, 0.28) > \mu_c$,
showing the expected linear behavior  Eq. (3.4). The moments
themselves as a function of $\mu$ 
for $p=1$ (the smoothest) to $p=5$ (right panel),
show a sharp change
at $\mu = \mu_c$  leading to  the qualitative modifications of the
$\theta$ distribution function described in the text. Note also the different
behavior of even (black symbols) and odd (red) moments, respectively (right).}
\end{figure}

\section{Summary, and comments}

We have studied the $\theta$ distribution and related properties 
within chiral perturbation  theory and one dimensional QCD. The qualitative
features of the results are the same. 
In either case the baryons are omitted by fiat, and 
despite the absence of baryons there is a highly
nontrivial $\mu$ dependence of these distributions.

Our main observation concerns the  
non trivial changes of the $\theta$ distribution when the chemical
potential $\mu$ exceeds the quenched threshold for pion condensation 
$\mu_c = m_\pi/2$, it can be summarized as follows:

\begin{itemize}
\item $\mu < m_\pi/2$ , i.e. quark mass outside the eigenvalues: 
Gaussian $\theta$ distribution.

$$
{\langle\delta(\theta-\theta')\rangle = \frac{1}{\sqrt{\pi
    \Delta G_0}} \sum_{n=-\infty}^{\infty}e^{-(\theta+2\pi
    n)^2/\Delta G_0}.}
$$

\item $\mu > m_\pi/2$ , i.e. quark mass inside the eigenvalues: 
Lorentzian $\theta$ distribution.
$$
{\langle\delta(2\theta-2\theta')\rangle=
\frac{1}{\pi} \frac{\sinh(V L_B)}{\cosh(V L_B)-\cos(2\theta)}}. 
$$
\end{itemize}

We reiterate that $\mu_c=m_\pi/2$ is not a critical point of the full
theory \cite{Splittorff:2006vj}, 
and any sharp change at this unphysical threshold implies
almost automatically specific difficulties in numerical approaches
relying on extrapolation from low chemical potential values.  
 
Further comments concern the  validity of the central limit 
theorem: for $\mu < m_\pi/2$
the distribution is Gaussian, thus fulfilling the conditions of 
the central limit theorem, and in agreement with the behavior found
in lattice simulations. However, 
the Lorentzian shape of the distribution of the phase valid for larger
values of the chemical potential shows that one should not take for granted
that the conditions for
the central limit theorem are satisfied. 

The analytical results show that exponentially large cancellations
may take place when integrating over $\theta$, needed to measure
correctly the baryon number. The extreme tail contributes significantly
to the results. A small non Gaussian term in the tail
of the $\theta$-distribution therefore could be the dominant term after
integration over $\theta$. The precise form of this tail is of course 
difficult to access numerically.

In QCD in one Euclidean dimension  the same behavior has been observed 
by a direct evaluation of the involved partition functions, either analytically
or numerically. We have studied in detail the behavior of the moments
of the distribution. The change of the distribution from Gaussian to
Lorentzian is clearly seen in the sharp change of the moments at $\mu_c$.
In the same simple model, the distribution of the gluonic observables
can be studied numerically. The results for the distributions of the plaquette
-- to be reported elsewhere \cite{lsv3} --  
can be used to demonstrate the effectiveness of the Fodor-Katz 
re-weighting \cite{Fodor:2001pe}:
we have explicitly shown that the use of configurations generated
at high temperatures maximizes the overlap with a cold, dense target
ensemble.

We close with a few  general comments:
the validity of the Gaussian assumption
of the distributions are limited  $\mu < m_\pi/2$:
Hence, the point  $\mu=m_\pi/2$ looks  more dangerous after this
study: qualitative
changes of the $\theta$ distribution, discontinuity of moments.

This rather pessimistic outcome might be partially mitigated by two 
observations: First, the $\theta$ distribution might be particularly 
difficult -- experiments with plaquette or other observables in simpler models
might be useful and give a different insight.
Second, and perhaps more significantly, the results of this study
do not include baryons. The physical motivations for the success of
any practical method for circumventing the sign problem relies
on baryonic fluctuations \cite{im1,im2}.
One way to see this is to 
consider 
the average phase factor from Taylor expansion \cite{bf}:
\begin{equation}
\langle e^{2i\theta}\rangle_{1+1^*} = e^{L^3T(c_2 - c_2^I)\mu^2}
\end{equation}
or, perhaps more simply, its log-derivative:
\begin{equation}
\frac{\partial}{\partial \mu} \log 
\langle e^{2i\theta}\rangle_{1+1^*} = \frac{\partial}{\partial \mu} \log {Z_{1+1}}
- \frac{\partial}{\partial \mu} \log {Z_{1+1^*}} 
\propto (n_B (\mu) - n_I (\mu))
\end{equation}
which equals the off diagonal susceptibility,
and indicates an interplay between the sign problem, the temperature,
and the baryonic contributions \cite{GG}, \cite{baryosign}. 
Current and future work will address this issue \cite{lsv2,lsv3}.


\begin{thebibliography}{99}

\bibitem{rev09} Ph. de Forcrand, plenary talk at Lattice09, this Volume;
 M.~P.~Lombardo,  
  J.\ Phys.\ G {\bf 35} (2008) 104019


\bibitem{lsv1}
  M.~P.~Lombardo, K.~Splittorff and J.~J.~M.~Verbaarschot,
  Phys.\ Rev.\  D {\bf 80}, 054509 (2009)


\bibitem{Fodor:2001pe}
  Z.~Fodor and S.~D.~Katz,
  JHEP {\bf 0203} (2002) 014

\bibitem{bf}   C.~R.~Allton {\it et al.} ,
  Phys.\ Rev.\ D {\bf 71}, 054508 (2005)


\bibitem{im1}
M.~Alford, A.~Kapustin, and F.~Wilczek,  Phys.~Rev. {\bf D 59} (1999) 054502. 


\bibitem{im2} M.~P.~Lombardo,
  Nucl.\ Phys.\ Proc.\ Suppl.\  {\bf 83} (2000) 375

\bibitem{im3} 
  P.~de Forcrand and O.~Philipsen,
  Nucl.\ Phys.\  B {\bf 642} (2002) 290

\bibitem{im4}
  M.~D'Elia and M.~P.~Lombardo,
  Phys.\ Rev.\  D {\bf 67} (2003) 014505

\bibitem{dos1}
 J.~Ambjorn, K.~N.~Anagnostopoulos, J.~Nishimura and J.~J.~M.~Verbaarschot,
  JHEP {\bf 0210}, 062 (2002)

\bibitem{dos2}
  Z.~Fodor, S.~D.~Katz and C.~Schmidt,
  JHEP {\bf 0703}, 121 (2007)


 \bibitem{method} 
K.~Splittorff and J.~J.~M.~Verbaarschot,
  Phys.\ Rev.\  D {\bf 77}, 01451

\bibitem{Ejiri}
  S.~Ejiri,
  Phys.\ Rev.\  D {\bf 77}, 014508 (2008)

\bibitem{Ejiri1}
  S.~Ejiri {\it et al.}  [WHOT-QCD Collaboration],
  arXiv:0909.2121 [hep-lat]

\bibitem{oned1}       
  N. Bilic and K. Demeterfi, 
  Phys. Lett. B, {\bf 212} (1988) 83

\bibitem{oned2} L.~Ravagli and J.~J.~M.~Verbaarschot,
  Phys.\ Rev.\  D {\bf 76} (2007) 054506

 \bibitem{Splittorff:2006vj}
          K.~Splittorff,
         PoS {\bf LAT2006}, 023 (2006)

\bibitem{GG}
  R.~V.~Gavai and S.~Gupta,
  Phys.\ Rev.\ D {\bf 68}, 034506 (2003)

\bibitem{baryosign} 
  M.~D'Elia and F.~Sanfilippo,
  Phys.\ Rev.\  D {\bf 80} (2009) 014502


\bibitem{lsv2}  
M.~P.~Lombardo, K.~Splittorff and J.~J.~M.~Verbaarschot,
  arXiv:0910.5482 [hep-lat].

\bibitem{lsv3} M.~P.~Lombardo, K.~Splittorff and J.~J.~M.~Verbaarschot,
In preparation.

\end{thebibliography}
\end{document}